\documentclass{emulateapj}

\usepackage{amssymb}
\usepackage{amsmath}

\newcommand{\rhoi}{\ensuremath{\bar{\rho}_{i}}}
\newcommand{\Ppi}{\ensuremath{\bar{P}_{i}}}

\newcommand{\tn}{\ensuremath{t^{n}}}
\newcommand{\tnp}{\ensuremath{t^{n+1}}}
\newcommand{\tcool}{\ensuremath{t_{\rm cool}}}

\newcommand{\Ti}{\ensuremath{\bar{T}_{i}}}
\newcommand{\Tin}{\ensuremath{\bar{T}_{i}^{n}}}
\newcommand{\Tinp}{\ensuremath{\bar{T}_{i}^{n+1}}}
\newcommand{\Tinpex}{\ensuremath{\bar{T}_{i,{\rm EI}}^{n+1}}}
\newcommand{\Tinh}{\ensuremath{\bar{T}_{i}^{n+1/2}}}

\newcommand{\Tref}{\ensuremath{T_{\rm ref}}}

\newcommand{\Tk}{\ensuremath{T_{k}}}
\newcommand{\Tkp}{\ensuremath{T_{k+1}}}

\newcommand{\TN}{\ensuremath{T_{N}}}
\newcommand{\TNm}{\ensuremath{T_{N-1}}}

\newcommand{\Lambdak}{\ensuremath{\Lambda_{k}}}
\newcommand{\Lambdaref}{\ensuremath{\Lambda_{\rm ref}}}
\newcommand{\LambdaN}{\ensuremath{\Lambda_{N}}}
\newcommand{\LambdaNm}{\ensuremath{\Lambda_{N-1}}}
\newcommand{\alphak}{\ensuremath{\alpha_{k}}}

\newcommand{\alphaNm}{\ensuremath{\alpha_{N-1}}}

\newcommand{\TEF}{\ensuremath{Y}}
\newcommand{\TEFk}{\ensuremath{\TEF_{k}}}
\newcommand{\TEFkp}{\ensuremath{\TEF_{k+1}}}
\newcommand{\TEFN}{\ensuremath{\TEF_{N}}}

\newcommand{\disca}{\ensuremath{D_{a}}}
\newcommand{\discb}{\ensuremath{D_{b}}}

\newcommand{\nel}{\ensuremath{n_{\rm e}}}
\newcommand{\nH}{\ensuremath{n_{\rm H}}}
\newcommand{\muel}{\ensuremath{\mu_{\rm e}}}
\newcommand{\muH}{\ensuremath{\mu_{\rm H}}}

\newcommand{\amu}{\ensuremath{{\rm u}}}

\newcommand{\Delt}{\ensuremath{\Delta t}}
\newcommand{\Delx}{\ensuremath{\Delta x}}

\newcommand{\rhoup}{\ensuremath{\rho_{\rm in}}}
\newcommand{\Tup}{\ensuremath{T_{\rm in}}}

\newcommand{\mach}{\ensuremath{\mathcal{M_{\rm in}}}}
\newcommand{\cmax}{\ensuremath{c_{\rm max}}}

\newcommand{\error}{\ensuremath{\varepsilon}}
\newcommand{\extime}{\ensuremath{\tau}}

\newcommand{\diff}{\ensuremath{{\rm d}}}

\begin{document}


\title{An Exact Integration Scheme for Radiative Cooling in Hydrodynamical Simulations}
 \shorttitle{An Exact Integration Scheme for Radiative Cooling in Hydrodynamical Simulations}


\author{R. H. D. Townsend}
\affil{Department of Astronomy,
       University of Wisconsin-Madison,
       5534 Sterling Hall, 475 N. Charter Street,
       Madison, WI 53706, USA}
\email{townsend@astro.wisc.edu}

\shortauthors{R. H. D. Townsend}


\begin{abstract}
A new scheme for incorporating radiative cooling in hydrodynamical
codes is presented, centered around exact integration of the governing
semi-discrete cooling equation. Using benchmark calculations based on
the cooling downstream of a radiative shock, I demonstrate that the
new scheme outperforms traditional explicit and implicit approaches in
terms of accuracy, while remaining competitive in terms of execution
speed.
\end{abstract}



\section{Introduction} \label{sec:intro}

Cooling by optically thin radiative emission plays an important role
in many differing types of astrophysical flow. This is especially true
of radiative shocks, which arise in a wide variety of contexts
\citep[e.g.,][and references therein]{StrBlo1995,Mig2005}. In this
paper, I present a new scheme for incorporating radiative cooling in
hydrodynamical codes, centered around exact integration of the
governing cooling equation.

To lay the necessary groundwork for the subsequent discussion,
Section~\ref{sec:cool} presents a derivation of the semi- and fully
discrete forms for the cooling equation. Section~\ref{sec:solve} then
reviews various schemes used to solve this equation, culminating with
the introduction of the new exact integration (EI) scheme. These
schemes are benchmarked in Section~\ref{sec:bench} to explore the
relative trade-offs between accuracy and execution speed, and I
conclude with brief remarks in Section~\ref{sec:conclude}.

\section{The Cooling Equation} \label{sec:cool}

The hydrodynamical equation of energy conservation for an ideal gas
can be written as
\begin{equation} \label{eqn:conserve}
\frac{\diff P}{\diff t} - \frac{\gamma P}{\rho} \frac{\diff
  \rho}{\diff t} = -(\gamma-1) \nel \nH \Lambda(T)
\end{equation}
Here, $P$ is the pressure, $\rho$ the density, $T$ the temperature,
$\gamma$ the ratio of specific heats, \nel\ and \nH\ the electron and
hydrogen number densities, respectively, and $\diff/\diff t$ denotes
the Lagrangian (total) time derivative. The function $\Lambda(T)$
represents the electron cooling efficiency, and is typically obtained
in tabular form from detailed modeling \citep[see,
  e.g.,][]{Ray1976,SutDop1993,GnaSte2007}. The dependence of $\Lambda$
on temperature alone is ultimately what makes the EI scheme possible,
but is also somewhat of an idealization of the underlying
physics. More-sophisticated treatments incorporate additional explicit
dependencies on ionization balance, by tracking a time-varying network
of ionic abundances \citep[see, e.g.,][]{Rag2000,Mig2007}. It is not
yet clear how the EI scheme might be extended to these treatments.

In Eulerian-based, finite-difference hydrodynamic codes, it is common
to implement energy conservation~(\ref{eqn:conserve}) using an
\emph{operator splitting} approach. The rate of pressure change is
divided into a component associated with adiabatic
expansion/contraction
\begin{equation} \label{eqn:dP-ad}
\left.\frac{\diff P}{\diff t}\right|_{\rm ad} =
\frac{\gamma P}{\rho} \frac{\diff
  \rho}{\diff t},
\end{equation}
and a component associated with radiative cooling,
\begin{equation} \label{eqn:dP-cool}
\left.\frac{\diff P}{\diff t}\right|_{\rm cool} = -(\gamma-1) \nel \nH
\Lambda(T).
\end{equation}
The pressure change due to the adiabatic component~(\ref{eqn:dP-ad})
is typically applied in the advection stage of the code, during which
the density and velocity are also updated in accordance with the mass
and momentum conservation equations\footnote{By maintaining an
  \emph{adiabatic} advection stage, it remains possible to use
  numerical schemes derived from Godunov's (\citeyear{God1959})
  characteristic-based approach, such as the popular Piecewise
  Parabolic Method (PPM) of \citet{ColWoo1984}}. The pressure change
due to the cooling component~(\ref{eqn:dP-cool}) is then applied in a
subsequent stage, during which the density is held constant
\citep[e.g.,][]{Mig2005}. The isochoric nature of this latter stage
does not preclude simulations of isobaric systems such as cooling
flows \citep[e.g.,][]{PetFab2006}; in these cases, the
components~(\ref{eqn:dP-ad},\ref{eqn:dP-cool}) are equal and opposite,
resulting in no net pressure variations.

The ideal gas law
\begin{equation} \label{eqn:ideal-gas}
P = n k T,
\end{equation}
is used to recast the cooling equation~(\ref{eqn:dP-cool}) in terms of
temperature; here, $n$ is the total number density of particles, and
$k$ is Boltzmann's constant. The mean molecular weight $\mu \equiv
\rho/n$ is assumed to remain constant; as \citep{GnaSte2007} argue,
this is a reasonable approximation for temperatures $\gtrsim
10^{4}\,{\rm K}$ \citep[although it can break down in circumstances
  where departures from ionization equilibrium are significant;
  see][]{Tes2008}. The cooling equation then becomes
\begin{equation} \label{eqn:cool}
\frac{\diff T}{\diff t} = 
- \frac{(\gamma - 1) \rho \mu}{k \muel \muH} \Lambda(T),
\end{equation}
where $\muel \equiv \rho/\nel$ and $\muH \equiv \rho/\nH$ are the
effective molecular weights per electron and per hydrogen atom/ion. In
the regime of full ionization, the molecular weights appearing in this
expression are given by
\begin{align}
\mu &= \frac{\amu}{2X + 3(1 - X - Z)/4 + Z/2} \\
\muel &= \frac{2\amu}{1 + X}, \\
\muH &= \frac{\amu}{X};
\end{align}
here, $X$ and $Z$ are the usual hydrogen and metal mass fractions, and
\amu\ is the atomic mass unit.

Implementing eqn.~(\ref{eqn:cool}) in a hydrodynamical code
requires discretization in space and perhaps also time. For
simplicity, I focus here on a zeroth-order finite-volume spatial
discretization\footnote{See \citet{StrBlo1995} for a demonstration of
  how a higher-order discretization might be constructed.}, which
leads to the semi-discrete cooling equation
\begin{equation} \label{eqn:semi-cool}
\frac{\diff \Ti}{\diff t} = - \frac{(\gamma - 1) \rhoi \mu}
{k \muel \muH} \Lambda(\Ti).
\end{equation}
Here, \rhoi\ is the volume-averaged density in the numerical zone with
integer index $i$, while \Ti\ is the corresponding zone temperature,
calculated from \rhoi\ and the volume-averaged pressure \Ppi\ using
the ideal-gas law~(\ref{eqn:ideal-gas}).

The following sections discuss various approaches to solving this
equation across a discrete time step \Delt.  The explicit
(\S\ref{ssec:solve-explicit}) and implicit
(\S\ref{ssec:solve-implicit}) schemes are all based on a fully
discrete cooling equation, derived from the semi-discrete
form~(\ref{eqn:semi-cool}) by replacing the temperature rate-of-change
with a finite difference:
\begin{equation} \label{eqn:fully-cool}
\frac{\Tinp - \Tin}{\Delt} = -\frac{(\gamma - 1)
  \rhoi \mu}{k \muel \muH} \Lambda(\Ti).
\end{equation}
The superscripts $n$ and $n+1$ indicate values at consecutive times
$\tn$ and $\tnp \equiv \tn + \Delt$.  Because the finite difference is
centered, this equation is second-order accurate in time. To evaluate
the cooling efficiency $\Lambda(\Ti)$ on the right-hand side, either
the initial or the updated temperature may be used; the choice
differentiates \emph{explicit} schemes from \emph{implicit} schemes.

\section{Solving the Cooling Equation} \label{sec:solve}

\subsection{Explicit Schemes} \label{ssec:solve-explicit}

\begin{figure*}
\epsscale{1.0} 
\plotone{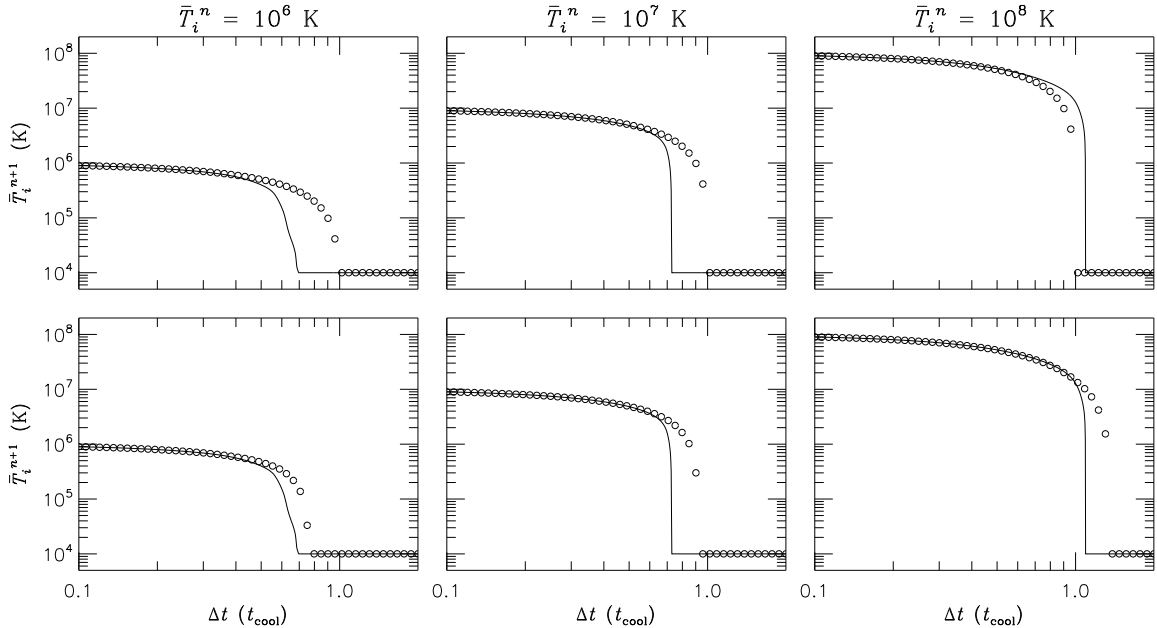}
\caption{The updated temperature \Tinp\ plotted as a function of time
  step \Delt\ (in units of the cooling time \tcool), for three
  differing choices (left-to-right) of initial temperature \Tin. The
  circles in the top (bottom) panels indicate values calculated using
  the first-order (second-order) explicit scheme. The solid lines show
  the corresponding exact solutions.} \label{fig:explicit-sol}
\end{figure*}

\begin{figure*}
\epsscale{1.0}
\plotone{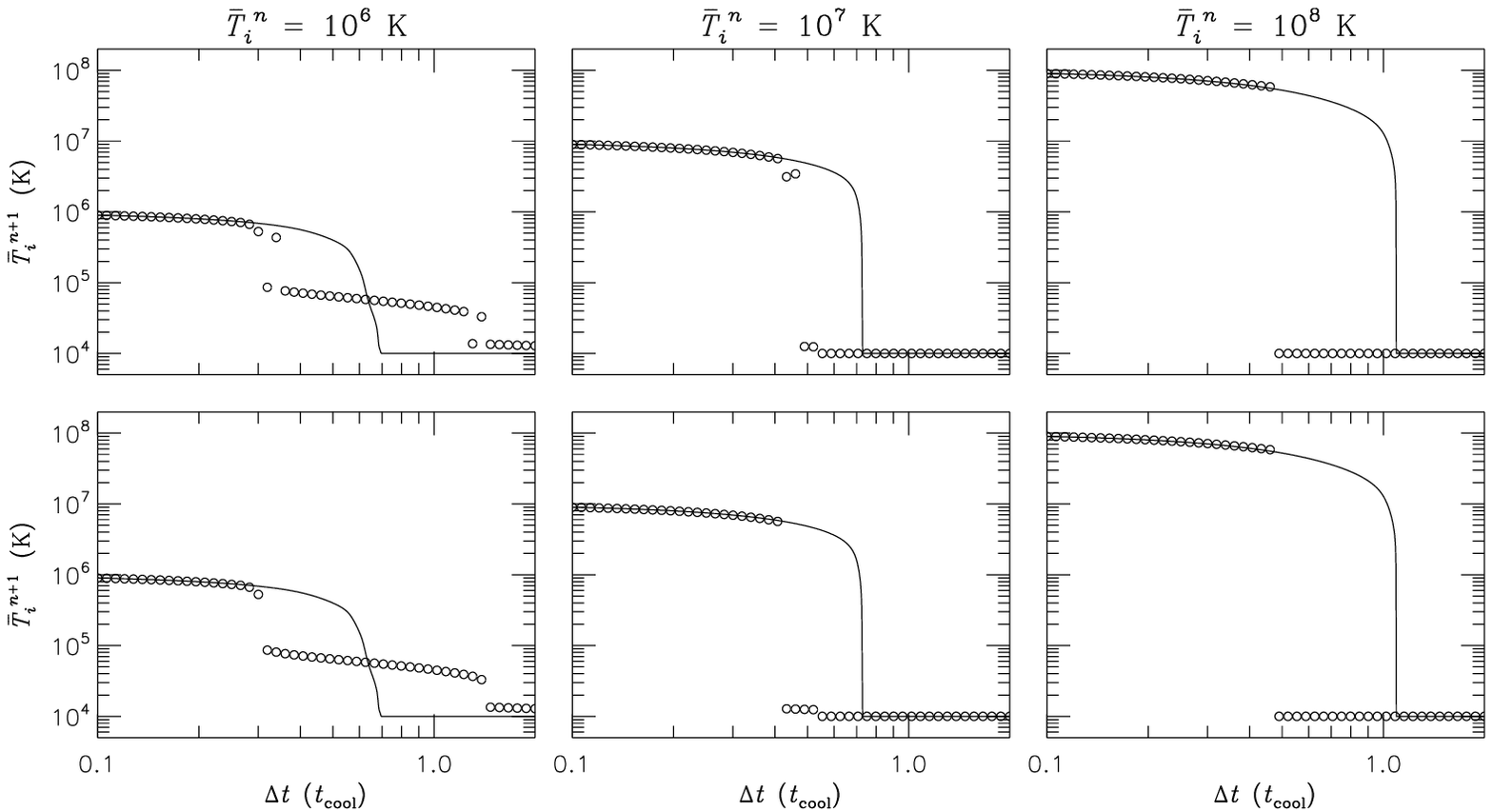}
\caption{As in Fig.~\ref{fig:explicit-sol}, except that the circles in
  the top (bottom) panels now indicate values calculated using the
  secant (Brent) first-order implicit scheme.} \label{fig:implicit-sol}
\end{figure*}

\begin{figure*}
\epsscale{1.0}
\plotone{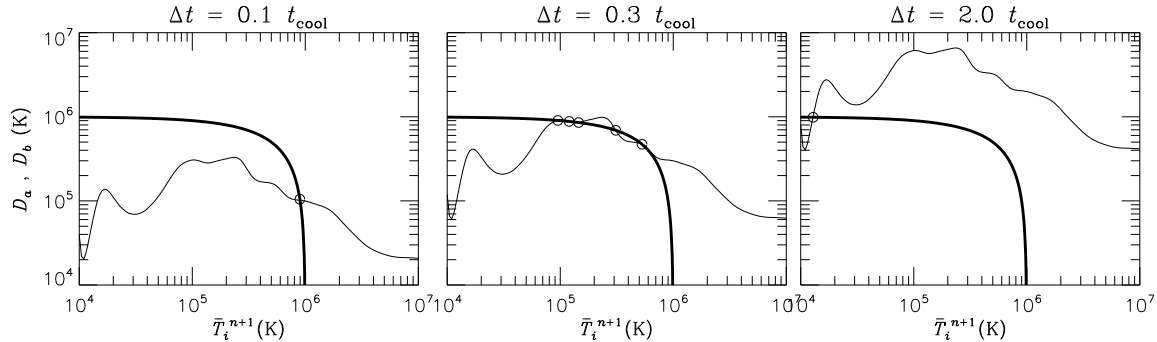}
\caption{The discriminants \disca\ (thick) and \discb\ (thin) plotted
  as a function of updated temperature \Tinp, for three differing
  choices (left-to-right) of the time step \Delt. Intersections
  between the two curves are highlighted by
  circles.} \label{fig:implicit-disc}
\end{figure*}

\begin{figure*}
\epsscale{1.0}
\plotone{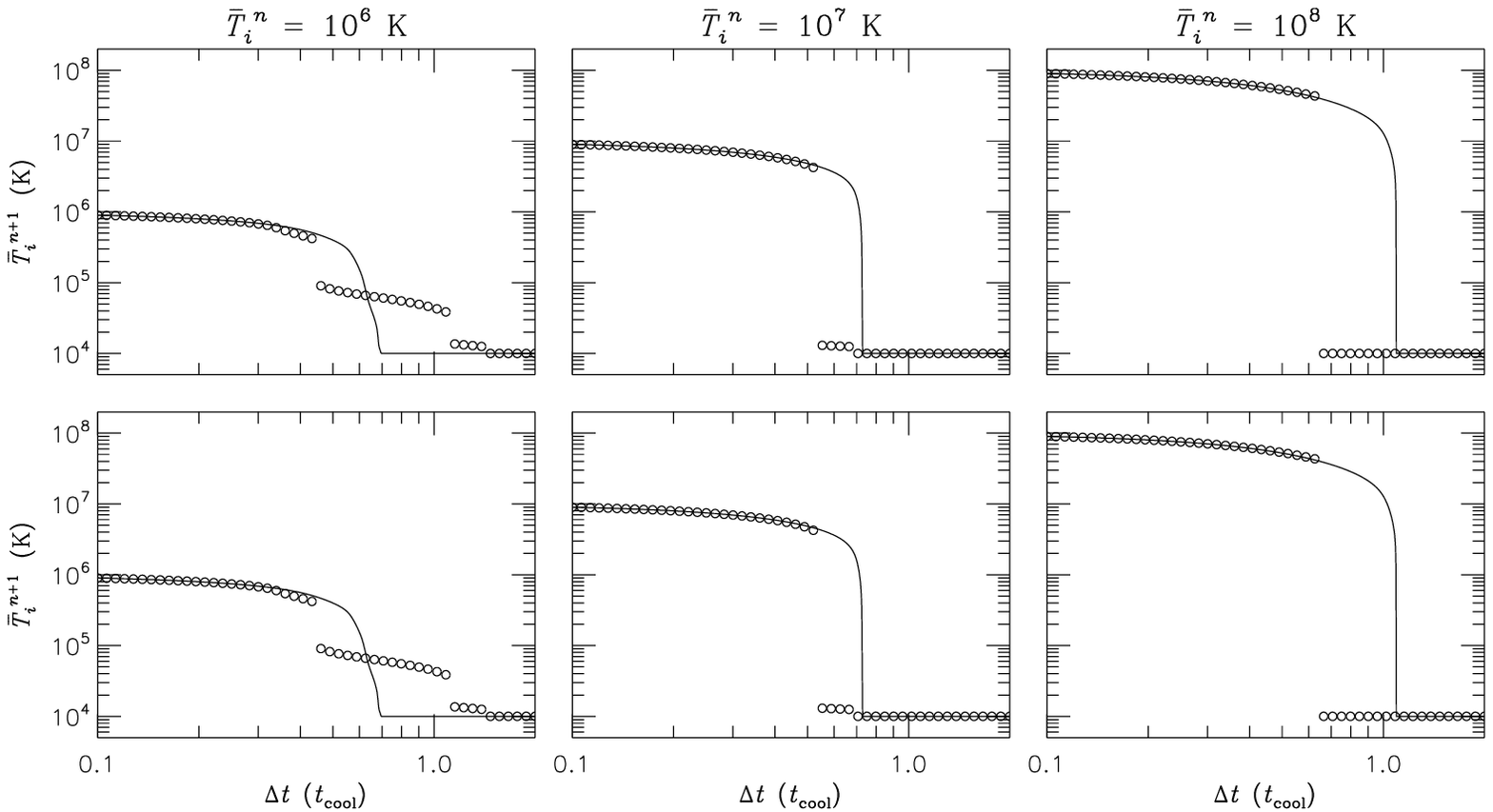}
\caption{As in Fig.~\ref{fig:explicit-sol}, except that the circles in
  the top (bottom) panels now indicate values calculated using the
  secant (Brent) second-order implicit scheme.} \label{fig:crank-sol}
\end{figure*}

In an explicit scheme, the cooling efficiency in
eqn.~(\ref{eqn:fully-cool}) is evaluated using the initial temperature
\Tin:
\begin{equation} \label{eqn:explicit-cool}
\frac{\Tinp - \Tin}{\Delt} = -\frac{(\gamma - 1)
  \rhoi \mu}{k \muel \muH} \Lambda(\Tin).
\end{equation}
This is now first-order accurate in time, because the right-hand side
is not centered in the interval $(\tn,\tnp)$. Solving for the updated
temperature,
\begin{equation} \label{eqn:explicit-sol}
\Tinp = \Tin \left[ 1 - \frac{\Delt}{\tcool} \right],
\end{equation}
where
\begin{equation}
\tcool \equiv \left[ \frac{(\gamma - 1) \rhoi \mu \Lambda(\Tin)}{k \muel \muH
  \Tin} \right]^{-1}
\end{equation}
is the single-point cooling time. This solution, together with the
ideal gas law~(\ref{eqn:ideal-gas}), allows the pressure in each zone
to be updated across the time step \Delt.

The behavior of the explicit scheme~(\ref{eqn:explicit-sol}) is
investigated by calculating the updated temperature \Tinp\ as a
function of time step, for three different choices of the initial
temperature: $\Tin = (10^{6}\,{\rm K}, 10^{7}\,{\rm K}, 10^{8}\,{\rm
  K})$. The cooling efficiency is obtained from a piecewise power-law
fit to the collisional ionization equilibrium (CIE) values tabulated
by \citet{GnaSte2007}. Because the tabulation is truncated at
$10^{4}\,{\rm K}$, this temperature is imposed as floor on \Tinp\ (in
an actual simulation, this floor temperature might correspond to the
reheating effects of a nearby star). A monatomic gas ($\gamma = 5/3$)
and solar abundances ($X=0.7$, $Z=0.02$) are assumed here and
throughout. The upper panels of Fig.~\ref{fig:explicit-sol} plot the
results from these calculations. By way of comparison, the panels also
show the exact solutions to the semi-discrete cooling
equation~(\ref{eqn:semi-cool}); Section~\ref{ssec:solve-exact} discusses how
these solutions are obtained.

For $\Delt$ approaching \tcool, the explicit scheme leads to updated
temperatures that depart quite significantly from the exact
values. Put simply, this is because the cooling efficiency is fixed at
its initial value $\Lambda(\Tin)$, rather than being allowed to evolve
in response to the cooling process. This difficulty can be avoided by
dividing the time step \Delt\ (typically set during the advection
stage by the Courant-Friedrichs-Lewy criterion; see \S\ref{sec:bench})
into a sequence of smaller sub-steps \citep[see, e.g.,][]{PleRoz1992},
and applying eqn.~(\ref{eqn:explicit-sol}) multiple
times. Alternatively, a higher-order temporal discretization of the
cooling equation is possible; for instance, a second-order Runge-Kutta
method has
\begin{align} \label{eqn:exp-sol-rk2}
\Tinh &= \Tin \left[ 1 - \frac{1}{2} \frac{\Delt}{\tcool} \right], \\
\Tinp &= \Tin \left[ 1 - \frac{\Lambda(\Tinh)}{\Lambda(\Tin)} \frac{\Delt}{\tcool}
\right].
\end{align}
The lower panels of Fig.~\ref{fig:explicit-sol} plot the updated
temperatures calculated using this second-order scheme. There is a
clear improvement over the first-order approach, and further
improvements can be gained by going to even-higher orders
\citep[e.g.,][]{Sut2003}. However, with each order added an additional
evaluation of the cooling efficiency is required; hence, the
computational costs necessarily escalate.

\subsection{Implicit Schemes} \label{ssec:solve-implicit}

To overcome the drawbacks of explicit schemes when $\Delt \gtrsim
\tcool$, a number of authors
\citep[e.g.,][]{StrBlo1995,Sto1997,Pit2004} instead opt for an
implicit scheme. The cooling efficiency in eqn.~(\ref{eqn:fully-cool})
is then evaluated using the updated temperature \Tinp:
\begin{equation}
\frac{\Tinp - \Tin}{\Delt} = -\frac{(\gamma - 1)
  \rhoi \mu}{k \muel \muH} \Lambda(\Tinp).
\end{equation}
The solution can be written in a standard form similar to the explicit
case (cf. eqn.~\ref{eqn:explicit-sol}),
\begin{equation} \label{eqn:implicit-sol}
\Tinp = \Tin \left[ 1 - \frac{\Lambda(\Tinp)}{\Lambda(\Tin)}
  \frac{\Delt}{\tcool} \right],
\end{equation}
but the appearance of \Tinp\ on the right-hand side means that this
equation must now be solved numerically, typically using a
root-finding algorithm.

Fig.~\ref{fig:implicit-sol} investigates the behavior of this implicit
scheme, plotting \Tinp\ as a function of \Delt\ for the same
parameters as in Fig.~\ref{fig:explicit-sol}. The upper panels use a
secant algorithm to solve eqn.~(\ref{eqn:implicit-sol}), with a
fallback to bisection when the most recent iterate for \Tinp\ falls
outside the bounds of the $\Lambda(T)$ tabulation. Conversely, the
lower panels use Brent's algorithm \citep{Pre1992}. In both cases,
solutions are iterated until the fractional change in \Tinp\ drops
below $10^{-4}$.

The figure reveals problems with the implicit schemes. For
instance, in the $\Tin = 10^6\,{\rm K}$ case, \Tinp\ is significantly
underestimated for $0.3\,\tcool \lesssim \Delt \lesssim 0.7\,\tcool$,
and overestimated for $\Delt \gtrsim 0.7\,\tcool$. Moreover, rather
than varying smoothly as \Delt\ is increased (as one might hope for a
stable scheme), \Tinp\ exhibits abrupt jumps. 

To explore the origin of these jumps, I introduce the twin
discriminants
\begin{equation} \label{eqn:disca}
\disca = \Tin - \Tinp
\end{equation}
and
\begin{equation} \label{eqn:discb}
\discb = \Tin \frac{\Lambda(\Tinp)}{\Lambda(\Tin)}
\frac{\Delt}{\tcool},
\end{equation}
such that the implicit equation~(\ref{eqn:implicit-sol}) corresponds
to the condition $\disca = \discb$. Fig.~\ref{fig:implicit-disc} plots
the discriminants together as a function of \Tinp, for $\Tin =
10^{6}\,{\rm K}$ and three choices of time step. The middle, $\Delt =
0.3\,\tcool$ panel shows that the curves intersect multiple times,
corresponding to multiple, distinct solutions (in this case, five) to
the implicit equation. Abrupt switching between these solutions, due
both to the convergence behavior of the particular root-finding
algorithm, and to the appearance or disappearance of solutions as
\Delt\ is varied, is responsible for the jumps seen in
Fig.~\ref{fig:implicit-sol}.

To underscore further that solution jumping is an intrinsic property
of implicit schemes, Fig.~\ref{fig:crank-sol} illustrates solutions of
the cooling equation using the Crank-Nicholson method,
\begin{equation}
\frac{\Tinp - \Tin}{\Delt} = -\frac{(\gamma - 1)
  \rhoi \mu}{k \muel \muH} \, \frac{\Lambda(\Tinp) + \Lambda(\Tin)}{2},
\end{equation}
which is now second-order implicit. In the standard form, this becomes
\begin{equation} \label{eqn:crank-sol}
\Tinp = \Tin \left[ 1 - \frac{\Lambda(\Tinp) + \Lambda(\Tin)}{2 \Lambda(\Tin)}
  \frac{\Delt}{\tcool} \right].
\end{equation}
While the data plotted in the figure differ from the first-order cases
shown in Fig.~\ref{fig:implicit-sol}, they still exhibit abrupt jumps
arising from the existence of multiple solutions --- although the
range of \Delt\ values over which jumping occurs is somewhat reduced.

In spite of these various issues, implicit schemes have proven popular
in the literature. This stems in part from their reputation for
stability; for instance, \citet{StrBlo1995} remark that their implicit
cooling scheme `\emph{is unconditionally stable}' \citep[words
  subsequently echoed by][]{Pit2004}; likewise, \citet{Sto1997} state
that their scheme `\emph{is stable even when the cooling time is much
  less than the dynamical time}'. However, this confidence appears
misplaced. While implicit schemes are stable when used to solve linear
equations \citep[as can be demonstrated through a von Neumann
  stability analysis; see, e.g.,][]{Pre1992}, this property does not
necessarily extend to \emph{non-linear} systems such as the
semi-discrete cooling equation~(\ref{eqn:semi-cool}). Moreover, even
if stability can be established for a given scheme, there are no
corresponding guarantees of accuracy or convergence --- and it is in
these latter capacities that the implicit schemes reviewed here fall
short. The jumping between solutions is particularly problematic,
because it can cause two neighboring zones with very similar initial
states to cool to quite different temperatures. This will establish a
strong pressure differential between the zones, in turn generating
spurious fluid flows and/or waves.

\subsection{The Exact Integration Scheme} \label{ssec:solve-exact}

The new cooling scheme introduced here avoids the various difficulties
outlined above, by going back to the semi-discrete cooling
equation~(\ref{eqn:semi-cool}) and solving it exactly. The equation is
first rearranged as
\begin{equation}
\frac{\diff \Ti}{\Lambda(\Ti)} = -\frac{(\gamma-1) \rhoi \mu}{k \muel \muH} \diff
t,
\end{equation}
and then integrated across a time step:
\begin{equation} \label{eqn:int-cool}
\int_{\Tin}^{\Tinp} \frac{\diff \Ti}{\Lambda(\Ti)} = -\frac{(\gamma-1)
  \rhoi \mu}{k \muel \muH} \Delt.
\end{equation}
The dimensionless `temporal evolution function' (TEF) is then
introduced as
\begin{equation} \label{eqn:TEF}
\TEF(T) = \frac{\Lambda(\Tref)}{\Tref} \int_{T}^{\Tref} \frac{\diff T'}{\Lambda(T')}
\end{equation}
where \Tref\ is an arbitrary reference temperature; the TEF represents
a normalized measure of the total time taken to cool from \Tref\ to
$T$. With this definition, the integrated cooling
equation~(\ref{eqn:int-cool}) becomes
\begin{equation}
\frac{\Tref}{\Tin}\frac{\Lambda(\Tin)}{\Lambda(\Tref)} [\TEF(\Tin) - \TEF(\Tinp)] = -\frac{\Delt}{\tcool},
\end{equation}
where \tcool\ has the same definition as before, and the consequent
solution is
\begin{equation} \label{eqn:exact-sol}
\Tinp = \TEF^{-1} \left[ \TEF(\Tin) + \frac{\Tin}{\Tref} \frac{\Lambda(\Tref)}{\Lambda(\Tin)} \frac{\Delt}{\tcool}
  \right].
\end{equation}
This result is exact, but requires construction of the TEF and its
inverse from $\Lambda(T)$. The Appendix presents analytic expressions
for $\TEF(T)$ and its inverse in the common cases where the cooling
efficiency is represented by a power law (\S\ref{app:TEF-power}) and a
piecewise power law (\S\ref{app:TEF-piece}). The cooling efficiencies
used in Figs.~\ref{fig:explicit-sol} and~\ref{fig:implicit-sol} fall
into the latter category, and the exact solutions plotted in these
figures are calculated using the EI scheme described here. Likewise,
the cooling efficiency assumed by \citet{Mig2005} falls into the
former category, and in fact his analytic cooling scheme (which
foreshadows the present paper) can be derived from this EI formalism.

\section{Benchmarks} \label{sec:bench}

The preceding sections (and in particular,
Figs.~\ref{fig:explicit-sol} and~\ref{fig:implicit-sol}) demonstrate
that both explicit and implicit schemes for solving the cooling
equation can become inaccurate as the time step approaches the cooling
time \tcool; in contrast, the EI scheme gives the exact solution for
any value of \Delt. However, an important caveat here is that the time
step is itself constrained by numerical considerations in the
advection stage. Efficiency dictates that \Delt\ be chosen as large as
possible (subject to accuracy requirements), but for stability reasons
it cannot exceed the limit established by the Courant-Friedrichs-Lewy
(CFL) criterion \citep[see, e.g.,][]{Lan1998}. 

To explore how the differing cooling schemes perform with a CFL-based
time step, I consider the problem of a steady,\footnote{In reality,
  radiative shocks are often time-variable due to the cooling
  instability discovered by \citet{Lan1981}; however, this variability
  is ignored here since the principal criterion is a
  \emph{well-defined} test system, even if it is somewhat idealized.}
1-dimensional radiative shock characterized by an upstream density
\rhoup, Mach number \mach\ and temperature \Tup. For various
combinations of these parameters (to be discussed below), each scheme
is benchmarked by repeating the following steps:

\begin{enumerate}

\item The run of density and pressure throughout the post-shock
  cooling region are calculated using the approach described by
  \citet[][their \S4.1]{StrBlo1995}. This region is bounded on the
  upstream side by the shock itself, and on the downstream side by the
  condition $T = \Tup$ (i.e., the gas has cooled back down to its
  initial temperature).

\item The cooling region is discretized into $N$ equal-sized zones;
  for each zone, the volume-averaged density \rhoi\ and pressure
  \Ppi\ are evaluated, and the corresponding temperature \Ti\ is
  calculated using the ideal-gas law~(\ref{eqn:ideal-gas}).

\item The CFL time step is calculated as $\Delt = \Delx /
  \cmax$, where \Delx\ is the spatial extent of the zones, and $\cmax
  \equiv \max(\sqrt{\gamma \Ppi/\rhoi})$ is the maximum value of the
  adiabatic sound speed over all zones composing the cooling region.

\item For each zone, the updated temperature \Tinp\ is evaluated using
  one of the cooling schemes discussed in the preceding sections. This
  step is repeated 5 times, and the average CPU execution time
  \extime\ (per zone, per repeat) is recorded.

\item The updated temperatures are compared with the exact values
  \Tinpex\ that result from using the EI scheme; the maximum relative
  error
  \begin{equation}
    \error = \max(|\Tinp - \Tinpex|/\Tinpex)
  \end{equation}
  is recorded.

\end{enumerate}

To cover a representative region of parameter space, I consider three
values $\mach = 3,10,100$ of the Mach number (corresponding to mild,
moderate and strong shocks), and three values $N = 1, 10, 100$ of the
zone count (corresponding to poor, moderate and good resolution of the
shocks). An upstream density $\rhoup = 10^{-15}\,{\rm g\, cm^{-3}}$ is
assumed throughout, but results do not depend at all on this
value. With these choices of parameters, and for each of the five
cooling schemes considered previously, Table~\ref{tab:bench} shows the
error \error\ and execution time \extime\ obtained by following the
steps above. All calculations were undertaken on a single core of an
Intel E5345 quad-core CPU running at 2.33 GHz.

\begin{deluxetable*}{lcccccc}
\tablecaption{Benchmark results}
\tablehead{
\multicolumn{1}{c}{} & \multicolumn{2}{c}{$N=1$} & \multicolumn{2}{c}{$N=10$} & \multicolumn{2}{c}{$N=100$} \\
Cooling Scheme & \error (\%) & \extime (ns) & \error (\%) & \extime (ns) & \error (\%) & \extime (ns)
}
\tablecolumns{7}
\tablewidth{5in}
\tabletypesize{\footnotesize}
\startdata
\cutinhead{$\mach = 3$}
$1^{\rm st}$-order explicit &  4.0 &   95 & 14.1 &  111 &  0.2 &  100\\
$2^{\rm nd}$-order explicit &  4.0 &  213 &  5.9 &  201 &  0.0 &  181\\
$1^{\rm st}$-order implicit (secant) & 26.1 &  844 &  4.2 &  484 &  0.2 &  362\\
$1^{\rm st}$-order implicit (Brent) & 26.1 &  903 &  4.2 &  787 &  0.2 &  562\\
$2^{\rm nd}$-order implicit (secant) & 22.0 &  931 &  2.4 &  837 &  0.0 &  612\\
$2^{\rm nd}$-order implicit (Brent) & 22.0 &  927 &  2.4 &  837 &  0.0 &  613\\
Exact &  0.0 &  213 &  0.0 &  192 &  0.0 &  173\\
\cutinhead{$\mach = 10$}
$1^{\rm st}$-order explicit & 30.8 &   94 &  7.6 &  112 &  0.3 &  101\\
$2^{\rm nd}$-order explicit &  3.0 &  215 &  0.4 &  204 &  0.0 &  183\\
$1^{\rm st}$-order implicit (secant) & 24.0 &  906 &  6.5 &  548 &  0.2 &  383\\
$1^{\rm st}$-order implicit (Brent) & 24.0 &  972 &  6.5 &  778 &  0.2 &  572\\
$2^{\rm nd}$-order implicit (secant) & 12.2 & 1230 &  1.3 &  875 &  0.0 &  640\\
$2^{\rm nd}$-order implicit (Brent) & 12.2 & 1248 &  1.3 &  874 &  0.0 &  639\\
Exact &  0.0 &  215 &  0.0 &  197 &  0.0 &  173\\
\cutinhead{$\mach = 100$}
$1^{\rm st}$-order explicit & 38.0 &   92 &  1.8 &  107 &  0.1 &   98\\
$2^{\rm nd}$-order explicit & 12.1 &  207 &  0.1 &  204 &  0.0 &  181\\
$1^{\rm st}$-order implicit (secant) & 99.9 &  266 &  1.2 &  523 &  0.1 &  387\\
$1^{\rm st}$-order implicit (Brent) & 99.9 &  258 &  1.2 &  725 &  0.1 &  544\\
$2^{\rm nd}$-order implicit (secant) & 99.9 &  315 &  0.3 &  796 &  0.0 &  598\\
$2^{\rm nd}$-order implicit (Brent) & 99.9 &  317 &  0.3 &  799 &  0.0 &  600\\
Exact &  0.0 &  214 &  0.0 &  195 &  0.0 &  169\\
\enddata  \label{tab:bench}
\end{deluxetable*}

The table reveals a general trend that the error decreases as the zone
count increases. When the shock is resolved by only a single zone, the
error tends to be large (with the obvious exception of the EI scheme,
for which $\error = 0$ always). For $N=10$, \error\ is below 10\% in
all but one case; and by $N=100$, it is below 1\% in all cases. To
explain this trend, the definition of the CFL time step is used to
write
\begin{equation}
\frac{\Delt}{\tcool} = \frac{\Delx}{\cmax \tcool}
\end{equation}
Because the flow downstream of the radiative shock is sub-sonic, it
follows that
\begin{equation}
\frac{\Delt}{\tcool} \lesssim \frac{\Delx}{v \tcool},
\end{equation}
where $v \lesssim \cmax$ is the typical flow velocity in the cooling
region. Recognizing that $v \tcool$ approximates the spatial extent of
this region, the corollary is that
\begin{equation}
\frac{\Delt}{\tcool} \lesssim \frac{1}{N}.
\end{equation}
Thus, the limit $N \gg 1$ implies that $\Delt \ll \tcool$, which
favors accurate cooling irrespective of the choice of scheme. Turning
this statement around, all of the cooling schemes apart from the EI
scheme tend to be inaccurate when the cooling region is poorly
resolved. Of course, this applies only when \Delt\ is tied solely to
the CFL time step. It is often desirable to place further constraints
on \Delt, over and beyond that given by the CFL criterion. For
instance, to improve the coupling between thermal and hydrodynamical
evolution, \Delt\ can be limited so that the anticipated
temperature/pressure change during cooling does not for any zone
exceed a specified fraction of its initial value. The price paid in
this approach is the greater number of steps that must be taken to
cross a given time interval.

Looking now at the relative performance of the differing schemes, the
\extime\ data in Table~\ref{tab:bench} reveal that the first-order
explicit scheme is the fastest, with an average execution time of
$\sim 100\,{\rm ns}$ per zone. The second-order explicit scheme and
the EI scheme are both only about factor of two slower than this, with
the latter slightly beating the former in all but two tests. The
implicit schemes are in every case the slowest, ranging from around
2.5 up to 12 times slower than the first-order explicit scheme.

\section{Concluding Remarks} \label{sec:conclude}

A criticism that might be leveled at the exact integration scheme is
that it requires the reciprocal of the cooling efficiency,
$1/\Lambda(T)$, be analytically integrable. In practice, this is
rarely an issue; the most common representations of $\Lambda(T)$ are
piecewise power-law or piecewise polynomial fits to detailed models,
both of which meet this restriction. In any case, it is always
possible to re-fit arbitrary cooling efficiency data with a conforming
representation.

The principal strengths of the EI scheme are twofold. On the one hand,
it produces exact solutions to the semi-discrete cooling
equation~(\ref{eqn:semi-cool}), irrespective of whether the time step
is small or large compared to the cooling time \tcool. On the other,
it remains very competitive in terms of execution speed, being only
two times slower than the (fastest, yet often inaccurate) first-order
explicit scheme. While more-sophisticated cooling treatments that
track ionic abundances \citep[e.g.,][]{Mig2007} will remain the
state-of-the-art in terms of physical fidelity, the strengths of the
EI scheme naturally recommend it as the cooling scheme of choice in
any hydrodynamical code where a simple, fast and robust treatment of
optically thin radiative losses is desired.


\acknowledgments

My thanks go to Stan Owocki, for many useful discussions that led to
the genesis of the paper, and to the anonymous referee for their very
helpful remarks. I moreover acknowledge support from NASA \emph{Long
  Term Space Astrophysics} grant NNG05GC36G and NSF grant AST-0507581.


\appendix

\section{Temporal Evolution Functions} \label{app:TEF}

\subsection{Power Law} \label{app:TEF-power}

I first consider the simple case of a power-law cooling function,
\begin{equation}
\Lambda(T) = \Lambdaref \left( \frac{T}{\Tref} \right)^{\alpha},
\end{equation}
where \Tref\ is the reference temperature introduced in
\S\ref{ssec:solve-exact}, and \Lambdaref\ and $\alpha$ are constant
coefficients. Substituting this into eqn.~(\ref{eqn:TEF}) leads to a
TEF
\begin{equation}
\TEF(T) =
\begin{cases}
\frac{1}{1-\alpha} \left[ 1 - \left( \frac{\Tref}{T} \right)^{\alpha-1}
  \right] & \qquad \alpha \neq 1, \\
\ln \left( \frac{\Tref}{T} \right) & \qquad \alpha = 1. \\
\end{cases}
\end{equation}
The corresponding inverse TEF is given by
\begin{equation}
\TEF^{-1}(\TEF) = 
\begin{cases}
\Tref \left[ 1 - (1 - \alpha) \TEF\right]^{1/(1-\alpha)} & \qquad \alpha
\neq 1, \\ \
\Tref \exp(-\TEF) & \qquad \alpha = 1. \\
\end{cases}
\end{equation}

\subsection{Piecewise Power Law} \label{app:TEF-piece}

More physically realistic cooling functions are often represented by
piecewise power-law fits to detailed models
\citep[e.g.,][]{WalFol1996,KimChe1997,CauKor2001,Tow2007}. I assume a
fit parametrization of the form
\begin{equation}
\Lambda(T) = \Lambdak \left( \frac{T}{\Tk} \right)^{\alphak} \qquad \Tk \leq T \leq \Tkp,
\end{equation}
for a set of $N-1$ temperature intervals $(\Tk,\Tkp)$
($k=1,2,\ldots,N-1$) and coefficient pairs
$\{\Lambdak,\alphak\}$. Substituting this into eqn.~(\ref{eqn:TEF}),
with a reference temperature chosen as $\Tref = \TN$, leads to the
piecewise TEF
\begin{equation}
\TEF(T) =  \TEFk + 
\begin{cases}
\frac{1}{1 - \alphak} \frac{\LambdaN}{\Lambdak} \frac{\Tk}{\TN} \left[ 
1 - \left( \frac{\Tk}{T} \right)^{\alphak-1} \right] & \qquad \alphak \neq 1\\
\frac{\LambdaN}{\Lambdak} \frac{\Tk}{\TN} \ln \left(
\frac{\Tk}{T} \right) & \qquad \alphak = 1\\
\end{cases}
\qquad \Tk \leq T \leq \Tkp,
\end{equation}
where $\LambdaN \equiv \LambdaNm (\TN/\TNm)^{\alphaNm}$. The
coefficients $\TEFk =\TEF(\Tk)$ are constants of integration; the
requirement that $\TEF(T)$ be continuous dictates that
\begin{equation}
\TEFk = \TEFkp - 
\begin{cases}
\frac{1}{1 - \alphak} \frac{\LambdaN}{\Lambdak} \frac{\Tk}{\TN} \left[ 
1 - \left( \frac{\Tk}{\Tkp} \right)^{\alphak-1} \right] & \qquad \alphak \neq 1,\\
\frac{\LambdaN}{\Lambdak} \frac{\Tk}{\TN} \ln \left(
\frac{\Tk}{\Tkp} \right) & \qquad \alphak = 1.\\
\end{cases}
\end{equation}
This recurrence can be started by noting that $\TEFN = \TEF(\Tref) =
0$ (cf.~eqn.~\ref{eqn:TEF}). The inverse TEF is given in each
$(\TEFk,\TEFkp)$ interval by
\begin{equation}
\TEF^{-1}(\TEF) = 
\begin{cases}
\Tk \left[ 1 - (1 - \alphak) \frac{\Lambdak}{\LambdaN} \frac{\TN}{\Tk}
(\TEF - \TEFk) \right]^{1/(1-\alphak)} & \qquad \alphak \neq 1 \\
\Tk \exp \left[ - \frac{\Lambdak}{\LambdaN} \frac{\TN}{\Tk} (\TEF -
  \TEFk) \right] & \qquad \alphak = 1 \\
\end{cases}
\qquad \TEFk \leq \TEF \leq \TEFkp.
\end{equation}


\bibliographystyle{aastex}
\bibliography{cooling}


\label{lastpage}

\end{document}